\documentclass[12pt]{article}

\usepackage{fullpage}
\usepackage{amsmath}
\usepackage{multirow}
\usepackage{mathrsfs}
\usepackage{amssymb}
\usepackage{epsfig}
\usepackage{subfig}
\usepackage{graphics}
\usepackage{bm}
\usepackage{graphics}
\usepackage{natbib}
\usepackage{xcolor}
\usepackage{algorithm}
\usepackage{algpseudocode}
\usepackage{float}
\usepackage{substr}
\usepackage{mathtools}
\usepackage{algorithm}
\usepackage{amsthm}
\usepackage{amsmath}
\usepackage{calc}  
\usepackage{enumitem} 
\usepackage[left=2.5cm,right=2.5cm,top=2.5cm,bottom=2.5cm]{geometry}

\usepackage{hyperref}
\hypersetup{colorlinks,linkcolor={blue},citecolor={blue},urlcolor={red}} 

\RequirePackage{graphicx,xcolor,ae,fancyvrb}
\RequirePackage[T1]{fontenc}
\IfFileExists{upquote.sty}{\RequirePackage{upquote}}{}
\IfFileExists{lmodern.sty}{\RequirePackage{lmodern}}{}

\DefineVerbatimEnvironment{Sinput}{Verbatim}{fontshape=sf}
\DefineVerbatimEnvironment{Soutput}{Verbatim}{}
\DefineVerbatimEnvironment{Scode}{Verbatim}{fontshape=sf}

\DefineVerbatimEnvironment{Code}{Verbatim}{}
\DefineVerbatimEnvironment{CodeInput}{Verbatim}{fontshape=sf}
\DefineVerbatimEnvironment{CodeOutput}{Verbatim}{}

\setkeys{Gin}{width=0.8\textwidth}

\graphicspath{{./Figs/}}

\theoremstyle{definition}
\newtheorem{definition}{Definition}
\newtheorem{theorem}{Theorem}
\newtheorem{remark}{Remark}
\newtheorem{proposition}{Proposition}
\newtheorem{algo}{Algorithm}

\def\T{{ \mathrm{\scriptscriptstyle T} }}

\DeclareMathOperator{\cor}{cor}
\DeclareMathOperator{\cov}{cov}

\renewcommand{\arraystretch}{1.5}

\begin{document}

\begin{center}
{\bf\large A New Multiple Correlation Coefficient without Specifying the Dependent Variable}
\end{center}

\begin{center}
  Kai Yang$^1$, Yuhong Zhou$^2$, Wei Xu$^2$, and Kirsten Beyer$^2$ \\
  $^1$Division of Biostatistics, Medical College of Wisconsin, Milwaukee, WI 53226 \\
  $^2$Division of Epidemiology, Medical College of Wisconsin, Milwaukee, WI 53226
\end{center}

\begin{abstract}

Multiple correlation is a fundamental concept with broad applications. The classical multiple 
correlation coefficient is developed to assess how strongly a dependent variable is associated 
with a linear combination of independent variables. To compute this coefficient, the dependent 
variable must be chosen in advance. In many applications, however, it is difficult and even 
infeasible to specify the dependent variable, especially when some variables of interest are equally 
important. To overcome this difficulty, we propose a new coefficient of multiple correlation which 
(a) does not require the specification of the dependent variable, (b) has a simple formula and
shares connections with the classical correlation coefficients, (c) consistently measures
the linear correlation between continuous variables, which is 0 if and only if variables are uncorrelated
and 1 if and only if one variable is a linear function of others, and (d) has an asymptotic distribution 
which can be used for hypothesis testing. We study the asymptotic behavior of 
the sample coefficient under mild regularity conditions. Given that the asymptotic bias of the sample
coefficient is not negligible when the data dimension and the sample size are comparable, 
we propose a bias-corrected estimator that consistently performs well in such cases. Moreover, we 
develop an efficient strategy 
for making inferences on multiple correlation based on either the limiting distribution 
or the resampling methods and the stochastic approximation Monte Carlo 
algorithm, depending on whether the regularity assumptions are valid or not. Theoretical and 
numerical studies demonstrate that our coefficient provides a useful tool for evaluating
 multiple correlation in practice. 

\end{abstract}


\section{Introduction}
\label{sec1}

The classical multiple correlation coefficient is designed for assessing the strength of the linear
relationship between a dependent variable and a set of independent variables. It provides a 
powerful tool for detecting linear associations in multivariate analysis and thus has been widely 
used in many fields. In some scenarios, however, it is hard and sometimes impossible to 
split the variables of interest into an dependent variable and independent variables. For example, in many applications, variables 
can be equally important, making it challenging to distinguish between the dependent variable and the 
independent variables. For such applications, measuring the intercorrelation among variables is 
important. So, there is a need for a new coefficient of multiple correlation
that captures these interrelationships without the need of
specifying the dependent variable, which is the focus of this paper.

There has been much discussion in the statistical literature about multiple correlation, with focus 
on the asymptotic distribution of the classical multiple correlation coefficient. For example, by using 
the characteristic geometric method, \cite{Fisher1928} first obtained the sampling distribution of the 
coefficient under the assumption of multivariate normality. In contrast, \cite{Wilks1932} derived the 
limiting distribution directly from the generalized product moment distribution that was discussed in 
\cite{Wishart1928}. Furthermore, \cite{Gurland1968}, \cite{Gurland1970} and \cite{Lee1971}
proposed various distribution approximation methods to simplify the expression of the sampling distribution 
function. There are some other methods proposed for studying the 
asymptotic distribution or the confidence limits of the classical multiple correlation coefficient. See, for instance, 
\cite{Grading1941}, \cite{Moran1950}, \cite{Williams1978}, \cite{Gurland1991}, \cite{Nandi2005},
\cite{Withers2015}. The aforementioned research works are typically based on the assumption of
multivariate normality. For variables not conforming to a single
multivariate normal distribution, \cite{Muirhead1982} 
derived the sampling distribution of the squared coefficient under the assumption of 
spherical symmetry, while \cite{Ali2002} obtained the asymptotic distribution of multiple correlation for normal 
mixture models. For moderately high dimensional data, \cite{Zheng2014} developed 
an asymptotic approximation method for the distribution of the squared coefficient under 
an independent component model. In this model, random components are assumed to be independent and identically
distributed with finite fourth moment, but they are not required to follow a normal distribution.

It should be pointed out that the classical multiple correlation coefficient is formulated based on 
multiple regression models \citep{Crocker1972}, in which one variable is treated as the dependent 
variable and others as the independent variables. So, it is not applicable when 
we are unwilling or unable to choose the dependent variable. To address the challenge, we propose 
a new coefficient for evaluating the linear correlation among continuous variables, which
has several good properties: (a) it does not require the specification of the dependent variable; (b) it is as 
simple as the classical multiple correlation coefficient, and it is equivalent to the absolute value of the Pearson's correlation 
coefficient when only two variables are involved; (c) it takes values in the interval from 0 to 1, which is 0 if and only 
if all variables are uncorrelated and 1 if and only if one variable can be expressed as a linear combination of 
the other variables; and (d) it has an asymptotic distribution for hypothesis testing.
We also provide a geometric interpretation of the newly proposed coefficient of multiple correlation.

Our newly proposed coefficient depends primarily on the determinant of the data correlation matrix 
instead of a multiple regression model, thus eliminating the need of selecting the dependent variable. 
In practice, the correlation matrix is typically unknown and needs to be estimated 
from a collected sample. Let $p$ and $n$ be the number of variables and the sample size, 
respectively. When $p<n$ and the ratio $p/n$ converges to a constant $c\in [0,1]$ as $n\to \infty$, we prove that the proposed new 
multiple correlation coefficient can be consistently estimated. 
A natural estimator is the sample multiple correlation coefficient, which can be obtained 
after replacing the population correlation matrix by the sample correlation matrix in the coefficient's definition.
Under the two assumptions that (i) the variables follow an independent
component model with independent but not necessarily identically distributed components and finite fourth moments and (ii) 
$p<n$ and $p/n\to c \in [0,1]$, 
we derive the asymptotic distribution of the sample correlation coefficient and thus obtain the 
asymptotic confidence interval and 
the analytical $p$-value for hypothesis testing. Observe that the asymptotic bias of the sample
correlation coefficient does not converge to 0 as $n\to \infty$ if $p/n$ does not converge to 0. We develop a new estimator of
the multiple correlation coefficient that is asymptotically unbiased in scenarios when $p<n$ and $p/n\to c\in [0,1]$. 
Finally, consider that the assumption of independent
component model with finite fourth moments
may not be satisfied in some applications. In such cases, we suggest computing the confidence limits and 
the $p$-value by the resampling-based methods such as the permutation and bootstrap procedures. To alleviate 
the computational obstacle in the resampling-based procedures for estimating small $p$-values, we design an 
efficient updating proposal distribution and incorporate it into the stochastic approximation Monte Carlo algorithm 
\citep{Liang2007, Yu2011} to reduce the number of the required resampling replications, such that the already very
popular resampling-based methods would become computationally feasible. 

The remainder of the paper is organized as follows. In Section \ref{sec2}, the proposed multiple
correlation coefficient is described in detail, along with the theoretical and computational methods for 
estimating the new coefficient and drawing statistical inference on it. In Section \ref{sec3},
the numerical performance of the proposed inference methods is evaluated using Monte Carlo simulations under several different
scenarios. In Section \ref{sec4}, the newly developed correlation coefficient is applied to an environmental racism dataset to assess 
the interrelationship among three environmental and racism variables. Some concluding
remarks are given in Section \ref{sec5}. Proofs of the theoretical results are provided in a supplementary file.

\section{Multiple Correlation Coefficient}
\label{sec2}

Our proposed multiple correlation coefficient is described in four parts in this section. 
The classical multiple correlation coefficient that motivates the development of a multiple correlation coefficient without specifying
the dependent variable is discussed 
in Subsection \ref{sec21}. The proposed correlation coefficient based on the data correlation matrix is described in 
Subsection \ref{sec22}. Some theoretical results about the proposed correlation coefficient are derived under mild regularity conditions and presented in 
Subsection \ref{sec23}. An efficient algorithm for drawing inference on the proposed correlation coefficient when the regularity assumptions are violated 
is given in Subsection \ref{sec24}.

\subsection{Review of the classical coefficient}
\label{sec21}

Suppose that $\mathbf{X}=(X_1,\ldots,X_p)^\mathrm{T}$ is a $p$-dimensional continuous random 
vector with mean $E(\mathbf{X})=\bm{\mu}$ and covariance matrix $\cov(\mathbf{X})=\bm{\Sigma}$. 
Let us partition the covariance matrix into
\[
\bm{\Sigma}=\left(\begin{array}{cc} \sigma_{11} &\bm{\sigma}_{12}^\mathrm{T}\\ 
\bm{\sigma}_{12}  & \bm{\Sigma}_{22}\\ \end{array} \right).
\]
Assume $\sigma_{11}\ne 0$ and $\bm{\Sigma}_{22}$ is nonsingular. Then,
the classical coefficient is defined as follows.
\begin{definition}
\label{def1}
The classical multiple correlation coefficient \citep{Muirhead1982} between the dependent variable $X_1$ and the
independent variables $X_2,\ldots,X_p$ is defined to be 
\begin{equation}
\label{equ1}
\rho=\max_{\beta_2,\ldots,\beta_p}\left\{\cor\left(X_1,\sum_{j=2}^p\beta_jX_j\right)\right\}
=\sqrt{\frac{\bm{\sigma}_{12}^\mathrm{T}\bm{\Sigma}_{22}^{-1}\bm{\sigma}_{12}}{\sigma_{11}}}.
\end{equation}
\end{definition}

In the literature, it has been well studied that, in the multiple linear model regressing $X_1$ on 
$X_2,\ldots,X_p$, $\rho^2$ is the proportion of the total variability in the dependent variable that can be 
explained by the independent variables \citep{Mudholkar2004}.

Let $\mathbf{x}_1,\ldots,\mathbf{x}_n$ be a sample of $\mathbf{X}$ and 
$\mathbf{x}_i=(x_{i1},\ldots,x_{ip})^\T$. From the sample,
the covariance matrix $\bm{\Sigma}$ can be estimated by the sample covariance matrix given below
\[
\hat{\bm{\Sigma}}=\left(\begin{array}{cc} \hat{\sigma}_{11} &\hat{\bm{\sigma}}_{12}^\mathrm{T}\\ 
\hat{\bm{\sigma}}_{12}  & \hat{\bm{\Sigma}}_{22}\\ \end{array} \right)
=\frac{1}{n-1}\sum_{i=1}^n\left(\mathbf{x}_i-\hat{\bm{\mu}}\right)
\left(\mathbf{x}_i-\hat{\bm{\mu}}\right)^\mathrm{T},
\] 
where $\hat{\bm{\mu}}=1/n\sum_{i=1}^n\mathbf{x}_i$ is the sample mean.
Here, we consider cases when $\hat{\sigma}_{11}\ne 0$ and $\hat{\bm{\Sigma}}_{22}$ is nonsingular. 
Then, it is natural to estimate the population multiple correlation coefficient $\rho$ 
by the classical sample multiple correlation coefficient, denoted as $R$, after replacing $\sigma_{11}$, $\bm{\sigma}_{12}$
and $\bm{\Sigma}_{22}$ in the definition (\ref{equ1}) by their respective estimates from the sample. In regression analysis, $R^2$ is 
referred to as the coefficient of determination, which is widely used to evaluate the goodness of fit of a linear 
regression model \citep{Casella2002}.

\subsection{The proposed new coefficient}
\label{sec22}

Let us closely examine the definition of the classical multiple correlation as provided in (\ref{equ1}). By matrix manipulation,
we have $\det(\bm{\Sigma})=\det(\bm{\Sigma}_{22})\left(\sigma_{11}-\bm{\sigma}_{12}^\mathrm{T}
\bm{\Sigma}_{22}^{-1}\bm{\sigma}_{12}\right)$. It follows that
\begin{equation}
\label{equ2}
\rho=\sqrt{\frac{\bm{\sigma}_{12}^\mathrm{T}\bm{\Sigma}_{22}^{-1}\bm{\sigma}_{12}}{\sigma_{11}}}
=\sqrt{\frac{\sigma_{11}-  \det(\bm{\Sigma})\left[\det(\bm{\Sigma}_{22})\right]^{-1}}{\sigma_{11}}}
=\sqrt{1-\frac{\det(\mathbf{V})}{\det(\mathbf{V}_{22})}},
\end{equation}
where $\mathbf{V}$ and $\mathbf{V}_{22}$ are the correlation matrices of $\mathbf{X}$ and $\mathbf{X}_2=(X_2,\ldots,X_p)^\T$,
respectively. Due to the presence of $\det(\mathbf{V}_{22})$ in Equation (\ref{equ2}),
it is evident that specifying the dependent variable $X_1$ (or the independent variables in $\mathbf{X}_{2}$) 
is necessary when calculating $\rho$.
In many applications, one may wonder if it is possible to define a simple coefficient of multiple correlation without specifying
the dependent variable. Such a coefficient is presented below. 

\begin{definition}
\label{def2}
Our newly proposed multiple correlation coefficient is defined as
\begin{equation}
\label{equ3}
\psi=\sqrt{1-[\det(\mathbf{V})]^{2/p}}.
\end{equation}
\end{definition}

It is clear from (\ref{equ3}) that $\psi$ does not contain the term $\det(\mathbf{V}_{22})$ in its definition, thereby eliminating the need
to select the dependent variable. Obviously, $\psi$ has a very simple formula, and it looks similar to the classical multiple 
correlation coefficient $\rho$. Additionally, when $p=2$, $\psi$ is the absolute value of the Pearson's correlation coefficient. 
Notice that $\mathbf{V}$ is the population
correlation matrix. Then, we have (i) $\psi\in [0,1]$, (ii) $\psi=0$ if and only if $\mathbf{V}=\mathbf{I}$,
where $\mathbf{I}$ is the identity matrix, and (iii) $\psi=1$ if and only if one variable is a linear combination of other variables.
Let 
\[
\hat{\mathbf{V}}=\mbox{diag}(\hat{\bm{\Sigma}})^{-1/2}\hat{\bm{\Sigma}}\mbox{diag}(\hat{\bm{\Sigma}})^{-1/2}
\]
be the sample correlation. The determinant of $\hat{\mathbf{V}}$
is the likelihood ratio test 
statistic for the hypothesis that variables
are independent but not necessarily identically distributed \citep{Jiang2013}. So, $\psi$ is powerful for testing the 
independence hypothesis in such cases. 

From the above discussion and the theoretical results in Subsection \ref{sec23}, the multiple correlation coefficient $\psi$ has
the properties (a)-(d) mentioned in both the abstract and Section \ref{sec1}. Here, we would like to point out that 
the formula (\ref{equ3}) is so simple that it is likely that there are many coefficients with similar characteristics, 
some of them possibly having better properties than $\psi$. However, to the best of our knowledge, $\psi$ is the first 
multiple correlation coefficient with the properties (a)-(d).

Our multiple correlation coefficient $\psi$ also has a geometric interpretation. Notice that $\det(\mathbf{V})^{2/p}$ 
represents the volume of a geometric shape called a scaled parallelepiped formed by the standardized variables \citep{Gover2010}.
A determinant closer to 0 indicates a high degree 
of linear dependence among the variables, resulting in a flattened parallelepiped, while a determinant closer to 1 implies greater independence, 
corresponding to a more voluminous parallelepiped.
Thus, $\psi^2=1-\det(\mathbf{V})^{2/p}$ can be interpreted as the remaining volume of a unit parallelepiped 
centered at the same point as the scaled parallelepiped formed by the standardized variables 
that is not occupied by the scaled parallelepiped.
For example, when $p=2$ and there are only two variables $X_1$ and $X_2$,  if the two variables are linearly dependent, it indicates that
all standardized data points lie along a straight line in a 2-dimensional space. In this case, the volume of the parallelepiped formed by $X_1$ and $X_2$
(i.e., the line) is 0, and the remaining volume of the unit parallelepiped not occupied by the line is $\psi^2=1$.

\begin{remark}
\label{rek1}
Let $\rho_j$ be the classical multiple correlation coefficient when $X_j$ is chosen as the dependent variable, for
$j=1,\ldots,p$. One may want to define a coefficient of multiple correlation 
as $\rho_{(a)}=\sum_{j=1}^p\rho_j/p$ or $\rho_{(m)}=\max(\rho_1,\ldots,\rho_p)$.
However, the property (c) does not hold for $\rho_{(a)}$ because
$\rho_{(a)}=1$ requires that all variables are linear combinations of the others.
This can obscure the identification of highly correlated variables, especially when the remaining variables are not correlated.
For example, consider a scenario when $X_1$ and $X_2$ are highly correlated, but the remaining variables
$X_3,\ldots,X_p$ are independent of each other and also independent of the pair $(X_1, X_2)$.
In such a case,  as the number of variables $p$ increases, $\rho_{(a)}$
could remain small, failing to accurately reflect the strong correlation between 
$X_1$ and $X_2$.
With respect to $\rho_{(m)}$, it remains constant when 
$\rho_j$ gets larger, as long as $\max(\rho_1,\ldots,\rho_p)\ne \rho_j$.
For example, let us consider the correlation matrix 
\[
\mathbf{V}=\begin{pmatrix}
1 \ & \alpha  \ & 0 \ & 0 \\
\alpha \ & \ 1 \ & 0 \ & 0 \\
0 \ & 0 \ & 1 \  &  \gamma\\
0 \ & 0 \ &  \gamma \ & 1\\
\end{pmatrix},
\]
where $0\leq \alpha\leq  \gamma\leq 1$ are positive numbers.
In such cases, $\rho_{(m)}$ always equals to $\gamma$ when $\alpha$ changes within $[0,\gamma]$.
As a comparison, $\psi$ would become larger when $\alpha\in [0,\gamma]$ increases. 
Thus, $\psi$ should be a more appropriate metric for assessing the multiple correlation among continuous variables.
Moreover, all the $p$ coefficients $\{\rho_j:j=1,\ldots,p\}$ need to be calculated to evaluate
$\rho_{(a)}$ and $\rho_{(m)}$, which is more computationally intensive, particularly in cases when 
$p$ is large. Additionally, the asymptotic distributions of 
$\rho_{(a)}$ and $\rho_{(m)}$ remain uncertain.
\end{remark}

\begin{remark}
\label{rek2}
Based on the data correlation matrix, one may define a coefficient of multiple correlation as 
$\psi^*=\sqrt{1-\det(\mathbf{V})}$, which also satisfies the properties (a)-(d) and even has a simpler expression than 
that of $\psi$. However, $\psi^*$ can be estimated consistently only if $p/n\to 0$ as $n\to \infty$
(cf., Proposition \ref{prop1} in Subsection \ref{sec23}). In addition, $\psi^*$ may converge to $1$
as $p\to \infty$ even if all pairwise correlations are relatively small. For example, when $p$ is even, let
\[
\mathbf{A}_\alpha=\begin{pmatrix}
1 & \alpha \\
\alpha & 1\\
\end{pmatrix},
\mathbf{V}=\mbox{diag}(\mathbf{A}_\alpha,\ldots,\mathbf{A}_\alpha),
\]
with $\alpha\in (0,1)$. Then,
$\psi^*=[1-(1-\alpha^2)^{p/2}]^{1/2}$ converges to 1 as $p\to \infty$. By constrast, 
$\psi$ equals to $\alpha$ in such cases. This explains why 
we choose to use $[\det(\mathbf{V})]^{2/p}$ instead of $\det(\mathbf{V})$ in the definition 
of the proposed multiple correlation coefficient.
\end{remark}

Next, we delve into the estimability of $\psi$. To this end, let us introduce a definition
regarding whether an unknown parameter can be consistently estimated.
\begin{definition}
\label{def3}
Suppose $\theta\in \Theta$ is an unknown parameter that belongs to a sample space $\Theta$. We say 
$\theta\in \Theta$ can be estimated consistently if the minimax risk for estimating 
$\theta$ satisfies
\[
\lim\limits_{n\to \infty}\inf\limits_{\tilde{\theta}}\sup_{\theta\in \Theta}E(\tilde{\theta}-\theta)^2=0,
\]
where the infimum is taken over all measurable estimators $\tilde{\theta}$ from a sample of size $n$.
\end{definition}

In the above definition, the minimax risk, which quantifies the performance of the best estimator
in the worst-case scenario, is a commonly used concept in statistical decision 
theory \citep{Donoho1990, Brown1991, Donoho1994, Rakhlin2017}. Next,
we demonstrate that $\psi$ can be estimated consistently in cases 
when $p<n$ and $p/n\to c\in [0,1]$ as $n\to \infty$. In such scenarios,
it is natural to estimate $\psi$ by the sample version of the proposed multiple correlation 
coefficient, denoted as $\hat{\psi}$ hereafter, which can be obtained by replacing 
$\mathbf{V}$ with $\hat{\mathbf{V}}$ in (\ref{equ3}). In the next subsection, 
we study the estimability and the asymptotic distribution of $\hat{\psi}$ 
when $p<n$ and $p/n \to c\in [0,1]$ as $n\to \infty$.

\subsection{Theoretical results}
\label{sec23}

In this part, we present some theoretical results about the newly proposed multiple correlation coefficient $\psi$, with the proofs given in 
the supplementary material. To do this, assume that $\mathbf{x}_i$ follows the independent component model 
\citep{Zheng2014, Lopes2019}
\begin{equation} 
\label{equ4}
\mathbf{x}_i=\bm{\Sigma}^{1/2}\mathbf{y}_i,\quad i=1,\ldots,n,
\end{equation}
where $\{\mathbf{y}_i:i=1,\ldots,n\}$ are $n$ samples of a $p$-dimensional random vector $\mathbf{Y}$ with 
independent but not necessarily identically distributed components $\{Y_j: j=1\ldots,p\}$. 
The model (\ref{equ4}) does not require that $\mathbf{y}_i$ follows a multivariate normal distribution or
the components have an identical distribution. Thus, it is flexible to use in practice.
Since $\psi$
is invariant under location changes, it is assumed without loss of generality that 
$E(\mathbf{Y})=\bm{0}$ and $\mbox{var}(\mathbf{Y})=\mathbf{I}$. Denote
\[
\mathbf{V}(\alpha)=(1-\alpha)\mathbf{I}+\alpha \bm{1}\bm{1}^\T,
\]
where $\alpha\in (-1,1)$ is a constant and $\bm{1}=(1,\ldots,1)^\T$.
For arbitrary $\gamma \in (0,1)$, define
\[
\mathbb{V}_\gamma=\left\{\mathbf{V}(\alpha): \alpha\in [0,\gamma]\right\}.
\]
Then, we have the following result.

\begin{proposition}
\label{prop1}
Assume that $\{\mathbf{x}_i: i=1,\ldots,n\}$ follow the independent
component model (\ref{equ4}) and $Y_1,\ldots,Y_p\sim N(0,1)$. 
Then, there exists a constant $C_\gamma>0$ such that,
for all $p< n$,
\begin{equation}
\label{equ5}
\inf\limits_{\tilde{\psi}^*}\sup_{\mathbf{V}\in \mathbb{V}_\gamma}E\left[\left\{\tilde{\psi}^*-\psi^*(\mathbf{V})\right\}^2\right]\geq C_\gamma
pn^{-1},
\end{equation}
where $\psi^*(\mathbf{V})$ is defined in Remark \ref{rek2} when $\mathbf{V}$ is the correlation matrix, and 
the infimum is taken over all measurable estimators $\tilde{\psi}^*$ from the sample $\{\mathbf{x}_i:i=1,\ldots,n\}$.
\end{proposition}

Proposition \ref{prop1} demonstrates that, when $p/n$ does not converge to 0, consistent
estimation of $\psi^*$ is impossible,
even under the assumption of multivariate normality with an exchangeable correlation matrix. 
This is one of the primary reasons why we refrain from using $\psi^*$ as a 
metric for assessing the multiple correlation. Next, we give an upper bound for the minimax risk for 
estimating the multiple correlation coefficient $\psi$ and demonstrate that $\psi$ can be
estimated consistently as long as $p<n$ and $p/n$ converges to a constant $c$ in $[0,1]$ .

\begin{theorem}
\label{thm1}
Let $\mathbb{V}$ be the collection of all $p\times p$ symmetric positive semidefinite
correlation matrices. Assume that $\{\mathbf{x}_i\}$ follow the model (\ref{equ4}) with 
$\max_{1\leq j\leq p}E(|Y_j|^4)\leq C_\nu$ for some constant $C_\nu>0$, $p<n$
and $p/n\to c\in [0,1]$. Then, there exists a constant $C_\psi>0$ such that
\begin{equation}
\label{equ6}
\inf\limits_{\tilde{\psi}}\sup_{\mathbf{V}\in \mathbb{V}}E\left[\left\{\tilde{\psi}-\psi(\mathbf{V})\right\}^2\right]\leq 
C_\psi \left(p^{-1/2}n^{-1/2}+p^{-1}n^{-1/6}\right),
\end{equation}
where $\psi(\mathbf{V})$ is defined in (\ref{equ3}) when the data correlation matrix is $\mathbf{V}$, and 
the infimum is taken over all measurable estimators $\tilde{\psi}$ from the sample $\{\mathbf{x}_i:i=1,\ldots,n\}$.
\end{theorem}

From Theorem \ref{thm1}, it is obvious that $\psi$ can be estimated consistently when $p<n$
and $p/n\to c\in [0,1]$ as $n\to \infty$. One natural estimator of $\psi$ is the sample coefficient $\hat{\psi}$, which can be
obtained by replacing the population correlation matrix $\mathbf{V}$ by the sample correlation matrix
$\hat{\mathbf{V}}$ in (\ref{equ3}). Next, we study the limiting distribution of $\hat{\psi}$ when $p<n$ and $p/n\to [0,1]$. 
To this so, we need to study the 
asymptotic behavior of the sample correlation matrix $\hat{\mathbf{V}}$. 

In the high-dimensional
data analysis literature, there has been much discussion about the asymptotic properties of 
$\hat{\mathbf{V}}$. For example, some research has been conducted to study the empirical 
distribution of the eigenvalues of $\hat{\mathbf{V}}$ or the quantity $\log[\det(\hat{\mathbf{V}})]$
when $\mathbf{V}=\mathbf{I}$ and $p/n\to c\in (0,1)$ \citep{Bao2012,Heiny2023}. 
Under the assumptions of finite fourth moment, $\mathbf{V}=\mathbf{I}$ and $p/n\to c\in (0,\infty)$,
\cite{Gao2017} established the central limit theorem for linear spectral statistics
of high-dimensional sample correlation matrices. 
The central limit theorem for linear spectral statistics has been extended to 
cases when the high-dimensional population follows an independent component model 
with identically distributed components
or an elliptical structure including some heavy-tailed distributions \citep{Yin2023}.
Another noteworthy work 
is that \cite{Jiang2019} demonstrated the central limit theorem for $\log[\det(\hat{\mathbf{V}})]$
even when $\mathbf{V}\ne \mathbf{I}$, under the assumptions that the ratio $p/n$ has
a nonzero limit, data
are generated from a multivariate normal distribution and the smallest eigenvalue of
the data correlation matrix $\mathbf{V}$ is greater than $0.5$.

Next, we derive the asymptotic distribution of
$\hat{\psi}$ when $p<n$, $p/n\to c\in [0,1]$ as $n\to \infty$ and data are assumed to follow 
the independent component model (\ref{equ4}) with finite fourth moments. 
Distinguished from the existing discussion in the literature, we
allow for cases when $p$ is fixed or $p$ increases slowly with $n$, such that the ratio $p/n\to 0$. Besides, we do 
not assume that the components $\{Y_i:i=1,\ldots,p\}$ are identically distributed or impose any restrictive
assumptions on the structure of the correlation matrix $\mathbf{V}$.

\begin{theorem}
\label{thm2}
Suppose that the assumptions in Theorem \ref{thm1} holds.
Then, $\hat{\psi}=1$ if $\psi=1$.
Let `$\ast$' be the Hadamard product, $\kappa=1/p\sum_{j=1}^pE(|Y_j|^4)$,
$\tau=\|\mathbf{V}^{1/2}\ast\mathbf{V}^{1/2}\|_F^2$, $\eta=\|\mathbf{V}-\mathbf{I}\|^2_F$, 
\begin{equation*}
\begin{split}
\delta_\nu=&2\left[1-{n}/p+3/(2p)\right]\log\left(1-{p}/n\right)-2+2/n+1/n
\left(\kappa-3\right)\left(\tau/p-1\right),\\
\sigma_\nu^2=&
-8[1/(p^2)\log(1-p/n)+1/(np)]+8 \eta/(np^2),\\
\end{split}
\end{equation*}
where $\|\cdot\|_F$ is the Frobenius matrix norm. Assume that $p<n$ and $p/n\to c\in [0,1]$. Then, 
\begin{equation}
\label{equ7}
\frac{\log(1-\hat{\psi}^2)-\log(1-\psi^2)-\delta_\nu}{\sigma_\nu}\xlongrightarrow{d} N(0,1),
\end{equation}
if $\psi\in [0,1)$,
where `$\xlongrightarrow{d}$' denotes convergence in distribution.
\end{theorem}

In the limiting distribution given in (\ref{equ7}), both the asymptotic bias $\delta_\nu$ and the asymptotic 
standard deviation $\sigma_\nu$ are unknown. To estimate them, the three parameters $\kappa$, $\tau$ and $\eta$ need to be 
estimated accurately. For $\tau$ and $\eta$, we suggest estimating them by
 \begin{equation*}
  \begin{split}
 \hat{\tau}=\mbox{tr}\left[\left(\hat{\mathbf{V}}^{1/2}\ast\hat{\mathbf{V}}^{1/2}\right)^2\right]-
\frac{1}{n} \left[\mbox{tr}\left(\hat{\mathbf{V}}^{1/2}\ast\hat{\mathbf{V}}^{1/2}\right)\right]^2, 
\hat{\eta}=\mbox{tr}\left[\left(\hat{\mathbf{V}}-\mathbf{I}\right)^2\right]-
\frac{1}{n}\left[ \mbox{tr}\left(\hat{\mathbf{V}}-\mathbf{I}\right)\right]^2.
 \end{split}
 \end{equation*}
Regarding $\kappa$, the estimator can be obtained by considering an estimating equation for $\kappa$
arising from the variance of a quadratic from \citep{Bai2010, Lopes2019}. To be specific,
under the assumption of the independent component model (\ref{equ4}) with
identically distributed components and finite fourth moment, we know that
\begin{equation}
\label{equ8}
\kappa=3+\frac{\nu-2\varsigma}{\omega},
\end{equation}
where 
$\nu=\mbox{var}(\sum_{j=1}^pX_j^2)$, $\varsigma=\|\bm{\Sigma}\|^2_F$,
$\omega=\sum_{j=1}^p\sigma_{jj}^2$ and $\sigma_{jj}=\mbox{var}(X_j)$.
The formula (\ref{equ8}) remains valid in our research setting, where
$Y_1,\ldots,Y_p$ may not follow an identical distribution but
their fourth moments are uniformly bounded. Consider that $E(\mathbf{X})=\bm{0}$.
Define 
\begin{equation*}
\begin{split}
\hat{\nu}=\frac{1}{n-1}\sum_{i=1}^n\left[\sum_{j=1}^px_{ij}^2-\frac{1}{n}\sum_{k=1}^n\sum_{j=1}^px_{kj}^2\right]^2,
\hat{\varsigma}=\mbox{tr}(\hat{\bm{\Sigma}}^2)-\frac{1}{n}\mbox{tr}(\hat{\bm{\Sigma}})^2,
\hat{\omega}=\sum_{j=1}^p\left(\frac{1}{n}\sum_{i=1}^nx_{ij}^2\right)^2.
 \end{split}
 \end{equation*}
Then, the corresponding estimator of $\kappa$ can be defined as follows
\begin{equation}
\label{equ9}
\hat{\kappa}=\max\left(3+\frac{\hat{\nu}-2\hat{\varsigma}}{\hat{\omega}},1\right).
\end{equation}
Note that $E(Y_j^2)=1$, for $j=1\ldots,p$. It follows that $\kappa$ is greater than or equal to 1. 
The continuous function $f(x)=\max(x,1)$ in (\ref{equ9}) enforces that $\hat{\kappa}\geq 1$.

After estimating the quantities $\tau$, $\eta$ and $\kappa$, we can obtain the estimators of ${\delta}_\nu$
and $\sigma_\nu$, denoted as $\hat{\delta}_\nu$ and $\hat{\sigma}_\nu$, by
substituting $\tau$, $\eta$ and $\kappa$ with their respective estimators in the definitions
of $\delta_\nu$ and $\sigma_\nu$ given in Theorem \ref{thm2}. 
The theorem below establishes the consistency of the two estimators 
$\hat{\delta}_\nu$ and $\hat{\sigma}_\nu$ under the same conditions are those
in Theorem \ref{thm1}.

\begin{theorem}
\label{thm3}
Under the assumptions in Theorem \ref{thm1}, when $p<n$ and $p/n\to c\in  [0,1]$, we have the following results:
as $n\to \infty$,
\begin{equation}
\label{equ10}
(\hat{\delta}_\nu-\delta_\nu)/\sigma_\nu \xlongrightarrow{p} 0,\quad 
\hat{\sigma}_\nu/\sigma_\nu \xlongrightarrow{p} 1,
\end{equation}
where `$\xlongrightarrow{p}$' denotes convergence in probability. Furthermore, if $\psi\in [0,1)$, we have
\begin{equation}
\label{equ11}
\frac{\log(1-\hat{\psi}^2)-\log(1-\psi^2)-\hat{\delta}_\nu}{\hat{\sigma}_\nu}\xlongrightarrow{d} N(0,1).
\end{equation}
\end{theorem}

It is important to notice from (\ref{equ7}), (\ref{equ10}) and (\ref{equ11}) that $\hat{\sigma}^2_\nu\to 0$ as $n\to \infty$ but $\hat{\delta}_\nu$ 
may not converge to $0$ when
the ratio $p/n$ does not converge to 0. Namely, the asymptotic bias of $\log(1-\hat{\psi}^2)$ (or $\hat{\psi}$) for estimating 
$\log(1-\psi^2)$ (or $\psi$) is not ignorable when $p$ and $n$ are comparable. In such cases,
from (\ref{equ11}), a bias-corrected estimator of 
$\log(1-\psi^2)$ can be defined naturally as $\log(1-\hat{\psi}^2)-\hat{\delta}_\nu$.
It follows that a bias-corrected estimator of $\psi$ is
\begin{equation}
\label{equ12}
\hat{\psi}_{bc}=\left[1-(1-\hat{\psi}^2)\exp(-\hat{\delta}_\nu)\right]^{1/2}.
\end{equation}
From (\ref{equ11}) and (\ref{equ12}), we have
\[
\frac{\log(1-\hat{\psi}_{bc}^2)-\log(1-\psi^2)}{\hat{\sigma}_\nu}\xlongrightarrow{d} N(0,1).
\]
Throughout the paper, the bias-corrected estimate $\hat{\psi}_{bc}$ is used as the final estimator of $\psi$.

Next, we briefly describe how to make statistical inference based on the asymptotic distributions
given in (\ref{equ7}) and (\ref{equ11}). 
Based on the estimators $\hat{\psi}_{bc}$ and $\hat{\sigma}_\nu$, 
the asymptotic $(1-\alpha)100\%$ confidence interval for $\psi$ has the lower and upper limits 
\begin{equation}
\label{equ13}
\begin{split}
L=\left[1-(1-\hat{\psi}_{bc}^2)\exp(z_{\alpha/2}\hat{\sigma}_\nu)\right]^{1/2},
U=\left[1-(1-\hat{\psi}_{bc}^2)\exp(-z_{\alpha/2}\hat{\sigma}_\nu)\right]^{1/2},
\end{split}
\end{equation}
where $z_{\alpha/2}$ is the upper $\alpha/2$ quantile of the standard normal distribution.
Another important topic is about the testing of the independence hypothesis $H_0: \psi=0$.
In the context of hypothesis testing, from (\ref{equ7}), we can simply choose
\begin{equation}
\label{equ14}
Z=\frac{\log(1-\hat{\psi}^2)-2\left[1-{n}/p+3/(2p)\right]\log\left(1-p/n\right)+2-2/n}
{\sqrt{-8[1/(p^2)\log(1-p/n)+1/(np)]}}
\end{equation}
as the test statistic, whose null distribution is $N(0,1)$.
Then, the $p$-value from the two-sided $z$-test is $2[1-\Phi(|Z|)]$, where
$\Phi(\cdot)$ is the distribution function of $N(0,1)$.

\subsection{Resampling-based statistical inference}
\label{sec24}

Theorem \ref{thm2} and \ref{thm3} are helpful for making statistical inference  
on $\psi$. However, they are derived under the assumption of independent component 
model with finite fourth moments. In some applications, this assumption may be violated and then
the asymptotic distribution in the two theorems would become unreliable. In such situations, the 
resampling-based procedures, such as permutation \citep{Good2005} and bootstrap \citep{Efron1994}, 
can be considered for assessing the $p$-value and for constructing the confidence interval.

Next, we describe how to evaluate the $p$-value for testing $H_0:\psi=0$
when the assumption of independent component model with finite fourth moments is 
invalid. In such cases, implementing the standard permutation
procedure \citep{Good2005} to estimate the $p$-value is straightforward. However, the  
permutation procedure requires repeated calculations 
on a very large number of permutation resamples to get 
a reliable estimator of the $p$-value, especially when the $p$-value is small \citep{Yu2011}.
To reduce the required number of permutation resamples, \cite{Yu2011} suggested an efficient $p$-value evaluation 
scheme based on the stochastic approximation Markov Chain Monte Carlo \citep{Liang2007}.

To use the $p$-value evaluation scheme suggested by \cite{Yu2011}, the updating proposal distribution 
must be specified properly. 
In this paper, we develop a permutation-based updating proposal designed to enhance
the efficiency of the stochastic approximation Monte Carlo algorithm when evaluating
the $p$-value for testing the independence hypothesis. The proposed
updating method is referred to as
$(n\varpi,p\varpi)$ out of $(n,p)$ updating proposal, where $\varpi\in (0,1]$ controls the
proportion of the data that needs to be updated in each iteration. This
updating proposal distribution is seamlessly incorporated into the stochastic approximation Monte Carlo 
scheme, and the resulting $p$-value evaluation algorithm is summarized below.

\begin{algo}
\label{al1}
$P$-value evaluation using 
the $(n\varpi,p\varpi)$ out of $(n,p)$ updating proposal.
\begin{itemize}
\item[(i)] 
Define $m$ subregions by distributing equally spaced cut points in $[0,|Z_1|]$ with
\begin{equation*}
\begin{split}
&E_1=\left\{\bm{x}:|Z(\bm{x})|\in\left[0,\frac{|Z_1|}{m-1}\right)\right\},
E_2=\left\{\bm{x}:|Z(\bm{x})|\in\left[\frac{|Z_1|}{m-1},\frac{2|Z_1|}{m-1}\right)\right\},\ldots,\\
&E_{m-1}=\left\{\bm{x}:|Z(\bm{x})|\in\left[\frac{(m-2)|Z_1|}{m-1},|Z_1|\right)\right\},
E_m=\left\{\bm{x}:|Z(\bm{x})|\in\left[|Z_1|,\infty\right)\right\},
\end{split}
\end{equation*}
where $Z_1$ and $Z(\bm{x})$ are the test statistics defined in (\ref{equ14}) that are calculated from the 
observed data $\bm{x}_1=(\mathbf{x}_1,\ldots,\mathbf{x}_n)^\T$ and any sample $\bm{x}$, and $m\geq 2$ is an integer.

\item[(ii)] For $i=1,\ldots,m$, let $\theta_{1i}=0$. From
$t=1$ to $T$, run the following two steps for $T$ iterations, where $T$ is a large positive integer.
\begin{itemize}
\item[(a)] The proposed updating proposal: given the current sample $\bm{x}_t$,  
randomly select $p^*$ integers from $\{1,\ldots,p\}$, where $p^*$ is the closest positive integer to
$p\varpi$ and $\varpi\in (0,1]$ is a pre-specified parameter. The selected integers are denoted as $\{i_1,\ldots,i_{p^*}\}$. Then, for $j=1,\ldots,p^*$,
randomly select $n^*$ observations in the $i_j$th column of $\bm{x}_t$ and permute them to replace the
selected data points, where
$n^*$ is the closest positive integer to $n\varpi$. Denote the generated data as $\bm{x}^*$.
\item[(b)] Calculate the ratio $r=\exp(\theta_{tJ(\bm{x}_t)}-\theta_{tJ(\bm{x}^*)})$,
where $J(\bm{x})=j$ if $\bm{x}\in E_j$.
With the probability of $\min(1,r)$, accept the proposed move and set $\bm{x}_{t+1}=\bm{x}^*$.
Otherwise, reject it and set $\bm{x}_{t+1}=\bm{x}_{t}$. Besides,
set $\theta_{t+1,i}=\theta_{ti}+\gamma_t[I(\bm{x}_{t+1}\in E_i)-1/m]$, where
$\gamma_{t}=\min(1,t_0/t)$, $t_0$ is pre-specified, and $I(\cdot)$ is the indicator function.
\end{itemize}

\item[(iii)] The final estimate of the $p$-value is 
\[
\frac{\exp(\theta_{Tm})(1/m+\Delta)}{\sum_{i=1}^m\exp(\theta_{Ti})(1/m+\Delta)} \quad \mbox{with}
\quad \Delta=\frac{m_0}{m(m-m_0)},
\]
where $m_0$ is the number of empty subregions that do not contain any samples.

\end{itemize}
\end{algo}

Our proposed $(n\varpi,p\varpi)$ out of $(n,p)$ updating approach takes advantage of the
null hypothesis of independence to permute partial samples for each
selected variable separately, with the proportions of variables and samples that we permute 
controlled by the parameter $\varpi$. Here, we choose to use a single proportion parameter for
both variables and samples to maintain a constant ratio, as we find
the ratio $p/n$ crucial in studying the distribution of $\hat{\psi}$. 
Through extensive simulation studies, we observe that the proposed 
updating approach performs well in practice, and we recommend
choosing $\varpi$ in the interval $[0.1,0.2]$. Notably, the relatively small $\varpi$ enhances computational 
efficiency compared to standard permutation methods that permute all variables and samples.
Moreover, the original formula for computing the ratio $r$ 
defined in Subsection 2.1 in \cite{Yu2011} is more complex. By using the proposed updating approach,
we simplify the formula into the one given in Step (b) in Algorithm \ref{al1}.
In addition to $\varpi$, there are some other
parameters that need to be specified in advance when implementing Algorithm \ref{al1}, including $m$, $t_0$ and $T$.
Regarding $m$ and $t_0$, we follow the practical guidelines in
\cite{Yu2011} and choose a relatively large number for $m$ (e.g., 100, 200 or 300) and
set $t_0$ to be 1,000 or 2,000. For the number of iterations $T$, it has been well studied in the
literature that a large value of $T$ should be used when the $p$-value is small. In this paper, we set
$T=10^6$ or $5\times 10^6$ and find these choices effective across all simulation examples, even when the 
$p$-value is as low as $10^{-5}$.

Finally, we briefly discuss how to construct confidence intervals for $\psi$ when the assumption 
of independent component model with finite fourth moments is violated.
When $p$ is small relative to $n$, we suggest using the
bootstrap procedure with replacement for computing the 
confidence limits. When $p$ and $n$ are comparable, however, it has been 
well studied in the literature that the traditional nonparametric bootstrap may give 
very poor performance as the ratio $p/n$ increases \citep{ElKaroui2018}.
The main difficulty is that, when $p$ is large, the $p$-dimensional distribution function of 
$\mathbf{X}$ cannot be approximated well by the empirical distribution function estimated from the 
collected data, unless the distribution has some low-dimensional structure. To overcome this difficulty, 
\cite{Lopes2019} proposed the spectral bootstrap for making inference on spectral statistics 
of the sample covariance matrix, under the assumptions that data follow an independent
component model with identically distributed components and 
finite fourth moment and $p/n\to (0,\infty)\setminus\{1\}$. Its idea 
is that the distributions of certain spectral statistics are mainly controlled by a small set of
parameters and thus the parametric bootstrap can be considered if these parameters can be 
estimated accurately. Recently, \cite{Wang2023} extended the parametric spectral bootstrap to situations
when data follow the elliptical model. Considering that $\hat{\psi}$ is a spectral statistic
of $\hat{\mathbf{V}}$, the spectral bootstrap for high-dimensional elliptical models can be considered 
when $p/n$ is large and data are generated from an elliptical model. 

\section{Simulation Studies}
\label{sec3}

In this part, we conduct simulation studies to investigate the numerical performance of the
asymptotic confidence interval (\ref{equ13}). We also compare the $p$-value evaluation approach using 
Algorithm \ref{al1} with the standard bootstrap and permutation 
procedures, in term of the estimation of small $p$-values. In the simulation examples, we consider two scenarios for the data dimension: (i) $p$ is 
fixed and $p=10$ or 20; and (ii) $p$ grows with $n$ and the ratio $p/n=0.2$ or 0.8, where
the sample size $n$ is chosen to be 200 or 500,
The observations $\{\mathbf{x}_i:i=1,\ldots,n\}$ are assumed 
to follow the independent component model
(\ref{equ4}). To generate data from the model, we need to specify the
covariance matrix $\bm{\Sigma}$, as well as the distributions of the components $Y_1,\ldots,Y_p$. 
For the covariance $\bm{\Sigma}$, we consider the following three cases:
\begin{itemize}
\item[] Case 1 (Autoregressive): $\bm{\Sigma}=(\phi^{|i-j|})_{i,j=1}^p$ where $\phi\in [-1,1]$ is a correlation coefficient.
\item[] Case 2 (Compound symmetry): $\bm{\Sigma}=(1-\phi)\mathbf{I}+\phi \bm{1}\bm{1}^\T $,where 
$\bm{1}=(1,\ldots,1)^\T$.
\item[] Case 3 (M-dependent): $\bm{\Sigma}=(\sigma_{ij})_{i,j=1}^p$, where 
$\sigma_{ii}=1$ and $\sigma_{ij}=\phi I(|i-j|=1)$ if $i\ne j$.
\end{itemize}
In Cases 1-3, the parameter $\phi$ is chosen such that the multiple correlation coefficient 
$\psi$ equals to 0.3, 0.6 or 0.9. Regarding $Y_1,\ldots,Y_p$, unless stated otherwise,
they are drawn in one of the following three ways and then standardized to have mean 0 and 
variance 1: 
(a) the variables follow the standard normal distribution and $\kappa=3$ in such cases;
(b) they follow the beta(6,6) distribution with $\kappa=2.6$;
(c) half of the variables follow the $t$-distribution
with 6 degrees of freedom, and the other half follow the beta(6,6) distribution, and the average of the fourth moments is
$\kappa=4.3$.
The beta distribution is an example of a platykurtic distribution, whereas the $t$-distribution
is leptokurtic. This simulation design enables us to assess the performance of the asymptotic confidence interval (\ref{equ13})
across various choices of the parameter $\kappa$ and in cases when the components of the independent component model (\ref{equ4}) are
independent but not identically distributed.

For each simulation setting, a sample of $n$ observations is generated from the independent component model.
Subsequently, the asymptotic $95\%$ confidence interval $(\ref{equ13})$ can be computed. The average coverage, representing the 
proportion of confidence intervals containing $\psi$, and the average length of the
intervals when $p$ is fixed at $10$ or $20$ are then calculated and reported in Table \ref{table1}, based on 10,000 
simulation replications. It is clear from the table that (i) our asymptotic confidence interval achieves reliable coverage probabilities 
in all simulation cases, even when the components do not have an identical distribution; and (ii) when $n$ gets larger, 
the asymptotic interval performs better, in the sense that the average coverage is closer to $95\%$
and the average interval length becomes shorter, which aligns with the theoretical results in Subsection \ref{sec23}. 

\begin{table}
\tabcolsep 4.2pt
\renewcommand{\arraystretch}{0.82}
\centering
\caption{Average coverage ($\%$) and length (in parentheses) of the analytical $95\%$ confidence
intervals for the multiple correlation coefficient $\psi$ when the data dimension $p$ is $10$ or $20$.} 
\begin{tabular}{ccc|cc|cc|cc}  \hline
            &         &           &\multicolumn{2}{c|}{$N(0,1)$}   & \multicolumn{2}{c|}{beta(6,6)}   & \multicolumn{2}{c}{$t$(6)+beta(6,6)} \\    
            &  $\psi$          &  $p$             & $n=200$ & $n=500$                & $n=200$ & $n=500$                  & $n=200$ & $n=500$ \\ \hline
            
Case 1& $0.3$ & $10$     & 97.3 (0.18) & 94.8 (0.09)     & 97.4 (0.18) & 94.7 (0.09)         & 97.3 (0.17) & 94.9 (0.09) \\
           &           & $20$     & 96.7 (0.15) & 95.6 (0.07)     & 97.0 (0.15) & 95.9 (0.07)         & 96.7 (0.15) & 95.8 (0.07) \\
           & $0.6$ & $10$     & 94.3 (0.10) & 94.9 (0.06)     & 94.6 (0.10) & 94.6 (0.06)         & 94.4 (0.10) & 95.0 (0.06) \\
           &           & $20$     & 94.7 (0.08) & 94.7 (0.04)     & 95.2 (0.08) & 95.1 (0.04)         & 95.5 (0.08) & 95.2 (0.04) \\
           & $0.9$ & $10$     & 94.2 (0.04) & 95.0 (0.03)     & 94.8 (0.04) & 94.8 (0.03)         & 94.6 (0.04) & 95.1 (0.03)  \\
           &           & $20$     & 94.7 (0.03) & 94.7 (0.02)     & 94.6 (0.03) & 94.8 (0.02)         & 95.1 (0.03) & 95.0 (0.02) \\ \hline
           
Case 2& $0.3$ & $10$     & 96.7 (0.20) & 94.1 (0.10)     & 96.8 (0.21) & 94.1 (0.10)         & 96.7 (0.20) & 94.3 (0.10) \\
           &           & $20$     & 97.7 (0.18) & 94.7 (0.09)     & 97.6 (0.18) & 94.6 (0.09)         & 97.7 (0.18) & 94.9 (0.09) \\
           & $0.6$ & $10$     & 93.8 (0.14) & 94.5 (0.08)     & 94.1 (0.14) & 94.8 (0.08)         & 93.9 (0.14) & 94.8 (0.08) \\
           &           & $20$     & 94.5 (0.12) & 94.6 (0.07)     & 93.9 (0.12) & 94.4 (0.07)         & 93.8 (0.12) & 94.7 (0.08) \\
           & $0.9$ & $10$     & 94.6 (0.06) & 94.9 (0.03)     & 94.8 (0.06) & 95.2 (0.03)         & 94.6 (0.06) & 95.2 (0.03)  \\
           &           & $20$     & 94.9 (0.05) & 95.1 (0.03)     & 94.4 (0.05) & 94.8 (0.03)         & 94.3 (0.05) & 95.0 (0.03) \\  \hline                      
           
Case 3& $0.3$ & $10$     & 96.7 (0.17) & 94.9 (0.08)     & 96.9 (0.17) & 94.9 (0.08)         & 96.9 (0.17) & 95.2 (0.08) \\
           &           & $20$     & 96.8 (0.15) & 95.9 (0.07)     & 97.0 (0.15) & 96.0 (0.07)         & 96.7 (0.15) & 95.8 (0.07) \\
           & $0.6$ & $10$     & 94.8 (0.08) & 94.9 (0.05)     & 95.2 (0.08) & 95.0 (0.05)         & 95.3 (0.08) & 95.0 (0.05) \\
           &           & $20$     & 95.6 (0.07) & 95.1 (0.04)     & 95.9 (0.07) & 95.4 (0.04)         & 95.6 (0.07) & 95.4 (0.04) \\
           & $0.9$ & $10$     & 94.8 (0.07) & 94.9 (0.04)     & 95.3 (0.07) & 95.0 (0.04)         & 95.3 (0.07) & 95.0 (0.04)  \\
           &           & $20$     & 95.6 (0.06) & 95.1 (0.03)     & 95.6 (0.06) & 95.4 (0.03)         & 95.5 (0.06) & 95.3 (0.03) \\  \hline
                                           
\end{tabular}
\label{table1}
\end{table}

We also explore scenarios where the data dimension $p$ increases with the sample size
$n$, with the ratio $q=p/n$ set at 0.2 or 0.8. The other setup
is the same as that of Table 1 in the main paper. The numerical results from 10,000 Monte
Carlo simulations are presented in 
Table \ref{table2}. From the table, similar conclusions can be drawn as those obtained from Table \ref{table1}. Notably,
by comparing the results in the two tables, we observe a slight decrease in the performance of the asymptotic 
confidence interval when $p$ grows with $n$. This is consistent with our intuition
because the estimation of $\mathbf{V}$ is less accurate when $p$ and $n$ are comparable, compared to cases
when $p$ is fixed and is small relative to $n$. This simulation example underscores the challenges of statistical inference in 
high dimensions.

\begin{table}[htp]
\tabcolsep 4.2pt
\renewcommand{\arraystretch}{0.82}
\centering
\caption{Average coverage ($\%$) and length (in parentheses) of the analytical $95\%$ confidence
intervals for the multiple correlation coefficient $\psi$ when the ratio $q=p/n$ is $0.2$ or $0.8$.} 
\begin{tabular}{ccc|cc|cc|cc}  \hline
            &         &          &\multicolumn{2}{c|}{$N(0,1)$}   & \multicolumn{2}{c|}{beta(6,6)}   & \multicolumn{2}{c}{$t$(6)+beta(6,6)} \\    
            &  $\psi$          &  $q$             & $n=200$ & $n=500$                & $n=200$ & $n=500$                  & $n=200$ & $n=500$ \\ \hline
            
Case 1& $0.3$ & $0.2$    & 97.9 (0.13) & 97.6 (0.05)     & 97.9 (0.13) & 97.9 (0.05)         & 97.3 (0.14) & 97.1 (0.05) \\
           &           & $0.8$    & 98.1 (0.15) & 98.1 (0.05)     & 98.6 (0.15) & 98.6 (0.05)         & 96.7 (0.15) & 96.2 (0.05) \\
           & $0.6$ & $0.2$    & 96.2 (0.06) & 95.9 (0.02)     & 96.1 (0.06) & 96.4 (0.02)         & 95.8 (0.06) & 95.5 (0.02) \\
           &           & $0.8$    & 97.0 (0.05) & 96.9 (0.02)     & 97.4 (0.05) & 97.2 (0.02)         & 95.9 (0.05) & 94.6 (0.02) \\
           & $0.9$ & $0.2$    & 94.1 (0.02) & 95.0 (0.01)     & 95.2 (0.02) & 95.2 (0.01)         & 94.6 (0.02) & 95.2 (0.01)  \\
           &           & $0.8$    & 95.8 (0.01) & 95.7 (0.01)     & 95.9 (0.01) & 95.7 (0.01)         & 95.4 (0.01) & 94.9 (0.01) \\ \hline
           
Case 2& $0.3$ & $0.2$    & 97.0 (0.16) & 95.9 (0.07)     & 97.0 (0.16) & 96.2 (0.07)         & 96.7 (0.16) & 94.2 (0.07) \\
           &           & $0.8$    & 98.4 (0.17) & 96.2 (0.07)     & 98.4 (0.17) & 96.5 (0.07)         & 98.4 (0.17) & 94.5 (0.07) \\
           & $0.6$ & $0.2$    & 93.7 (0.11) & 94.5 (0.06)     & 93.8 (0.11) & 94.8 (0.06)         & 93.9 (0.11) & 94.7 (0.06) \\
           &           & $0.8$    & 93.9 (0.10) & 94.1 (0.06)     & 93.9 (0.10) & 94.4 (0.06)         & 93.0 (0.10) & 94.3 (0.06) \\
           & $0.9$ & $0.2$    & 94.1 (0.05) & 94.7 (0.03)     & 93.8 (0.05) & 94.6 (0.03)         & 94.1 (0.05) & 94.7 (0.03)  \\
           &           & $0.8$    & 93.5 (0.05) & 94.4 (0.03)     & 93.5 (0.05) & 93.8 (0.03)         & 93.6 (0.05) & 95.5 (0.03) \\  \hline

Case 3& $0.3$ & $0.2$    & 98.0 (0.13) & 97.8 (0.05)     & 97.9 (0.13) & 96.9 (0.05)         & 97.3 (0.13) & 97.3 (0.05) \\
           &           & $0.8$    & 98.1 (0.14) & 97.8 (0.05)     & 98.6 (0.14) & 97.4 (0.05)         & 96.8 (0.15) & 95.8 (0.05) \\
           & $0.6$ & $0.2$    & 96.8 (0.05) & 96.3 (0.02)     & 96.8 (0.05) & 95.9 (0.02)         & 96.2 (0.05) & 95.7 (0.02) \\
           &           & $0.8$    & 97.3 (0.05) & 96.3 (0.02)     & 97.5 (0.05) & 96.7 (0.02)         & 96.3 (0.05) & 94.5 (0.02) \\
           & $0.9$ & $0.2$    & 96.7 (0.04) & 95.9 (0.02)     & 96.5 (0.04) & 95.9 (0.02)         & 96.2 (0.04) & 95.2 (0.02)  \\
           &           & $0.8$    & 97.2 (0.05) & 96.8 (0.02)     & 97.5 (0.05) & 96.5 (0.02)         & 96.3 (0.05) & 94.7 (0.02) \\  \hline
                                           
\end{tabular}
\label{table2}
\end{table}

\begin{figure}[htp]
\centering
\includegraphics[width=6.5in]{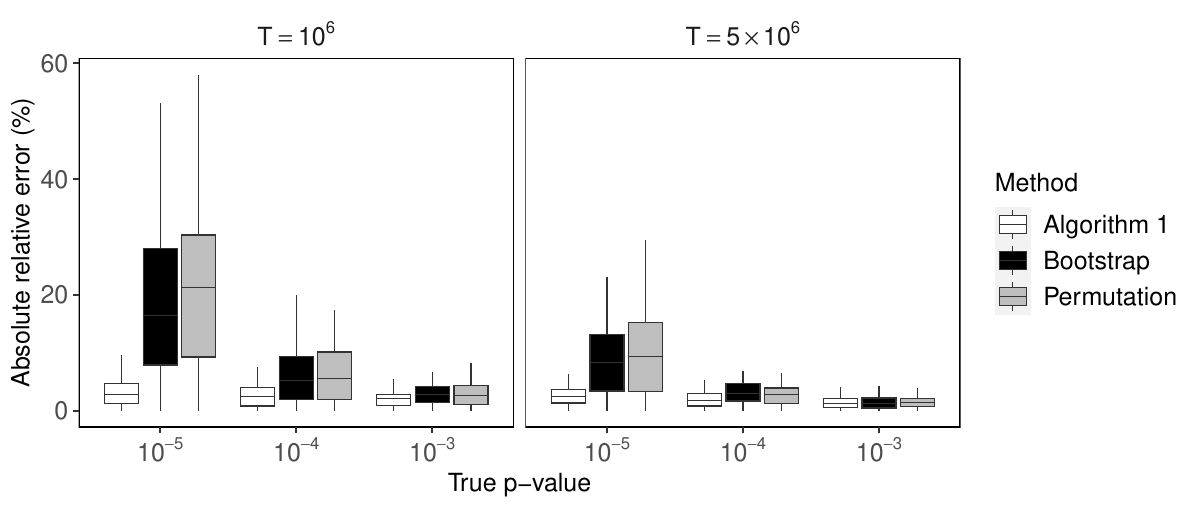} 
\caption{The boxplots of the average relative errors ($\%$) for the three $p$-value evaluation methods Bootstrap, 
Permutation and SAMC based on 100 replicated simulations.}
\label{fig1}
\end{figure}

In the following example, we demonstrate the efficacy of Algorithm \ref{al1} by comparing it with the standard bootstrap and permutation 
procedures for evaluating small $p$-values. Suppose that we use the test statistic $Z$ defined in (\ref{equ14})
to test the independence hypothesis. Let us focus on cases when $\mathbf{\Sigma}=\mathbf{I}$, $n=500$ and $p=100$
and simulate data from the independent component model (\ref{equ4}) in which the components follow the $t(4)$ distribution.
In such cases, the fourth moment of the components does not exist, making the asymptotic results in 
Theorems \ref{thm2} and \ref{thm3} inapplicable. Instead, we want to rely on a bootstrap or permutation
procedure to evaluate the significance level for a given observed test statistic value. In the bootstrap (or permutation) procedure,
under the null hypothesis, we sample with (or without) replacement from the data of the $j$th variable $X_j$, for $j=1,\ldots,p$,
to generate new data. In addition to the standard bootstrap and permutation procedures, we also apply
the Algorithm \ref{al1} to the simulated data. For Algorithm \ref{al1}, we choose $\varpi=0.2$, $m=300$, and $t_0=1,000$. 
In this example, the number of 
iterations $T$ is set to be $10^6$ or $5\times 10^6$ for all the three methods. As the true distribution of
$Z$ is unknown, we run $10^{9}$ Monte Carlo simulations and treat the empirical distribution 
of the $10^{9}$ test statistic values as the true distribution. To assess the numerical performance of the three methods,
we compute the average relative error, defined as $|\hat{p}_\nu-p_\nu|/p_{\nu}$,
where $p_\nu$ and $\hat{p}_\nu$ are the true and the estimated $p$-values from bootstrap, permutation or
Algorithm \ref{al1}, respectively. 
Figure \ref{fig1} summarizes the absolute relative errors of the
three methods for estimating the $p$-values of $10^{-5}$, $10^{-4}$ and $10^{-3}$
based on 100 Monte Carlo replications. From the figure, we can see that (i) Algorithm \ref{al1} has satisfactory performance with median
absolute relative errors less than $5\%$ in all cases considered; (ii) Algorithm \ref{al1} outperforms standard bootstrap and permutation procedures, 
especially when the 
$p$-value and $T$ are small; (iii) when $T$ and the $p$-value increase, all three methods produce
more accurate estimates; and (iv) bootstrap shows similar performance to that of permutation. 
This example confirms the benefit of using Algorithm \ref{al1} for evaluating small $p$-values
in cases when the assumption that data follow an independent component model with finite fourth
moment is violated.

\section{Analysis of Environmental Racism Data}
\label{sec4}

\begin{figure}[htb]
\centering
\includegraphics[width=6.5in]{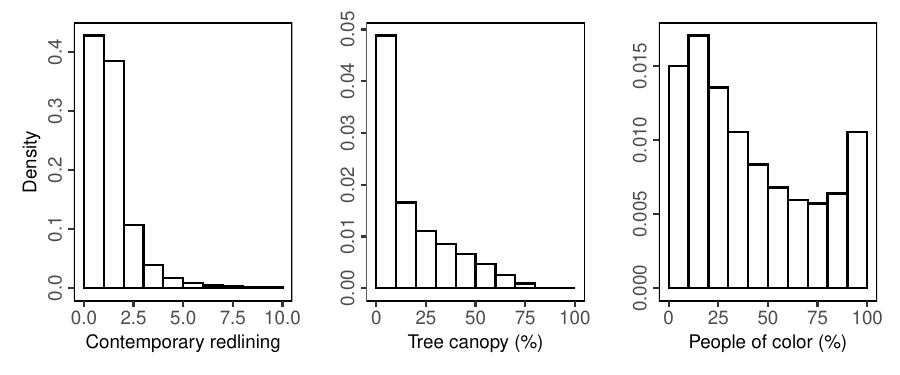} 
\caption{The histograms of the three variables: contemporary redlining, percentage of
tree canopy, and percentage of people of color.}
\label{fig2}
\end{figure}

We now present an application of the proposed multiple correlation coefficient to an environmental racism dataset.
This dataset contains $n=60,017$ observations of $p=3$ variables, including (i) contemporary redlining 
(a measure of place-based bias in mortgage lending, \cite{Beyer2016}), (ii) percentage of tree canopy, and (iii) 
percentage of people of color. The histograms of the three variables are presented in Figure \ref{fig2}.

This dataset originates from a study on environmental racism, which
aims to assess the strength of the interconnected 
relationship among three equally important variables including structural racism in contemporary mortgage lending, exposure to environmental benefits 
from tree canopy and concentration of populations of color, using a single correlation coefficient. 
This three-way correlation quantifies the extent to which individuals from minoritized racial and ethnic backgrounds 
reside in more disinvested neighborhoods with fewer environmental benefits, specifically from tree canopy.
It enables a deeper investigation into the magnitude and impact of environmental racism. 
To evaluate the three-way correlation,
our coefficient 
$\psi$ can be computed to summarize the interrelationship among the variables.

In addition to $\psi$, we also calculate the classical coefficient $\rho$.
When treating contemporary redlining, percentage of tree canopy and percentage of people of color
individually as the dependent variable, the resulting classical coefficients are 0.404, 0.377, and 0.523, respectively. 
Given the equal importance of these variables, it is challenging to determine which one of the three values to report. 
By contrast, there is no need to choose the dependent variable when computing $\psi$.
By using the formulas (\ref{equ12})-(\ref{equ14}), the bias-corrected estimator of the multiple
correlation coefficient $\psi$ is $\hat{\psi}_{bc}=0.442$, with the asymptotic $95\%$ 
confidence interval $(0.436, 0.447)$ and the $p$-value for testing $H_0:\psi=0$ less than $2.2\times 10^{-16}$.

\begin{figure}[htb]
\centering
\includegraphics[width=5in]{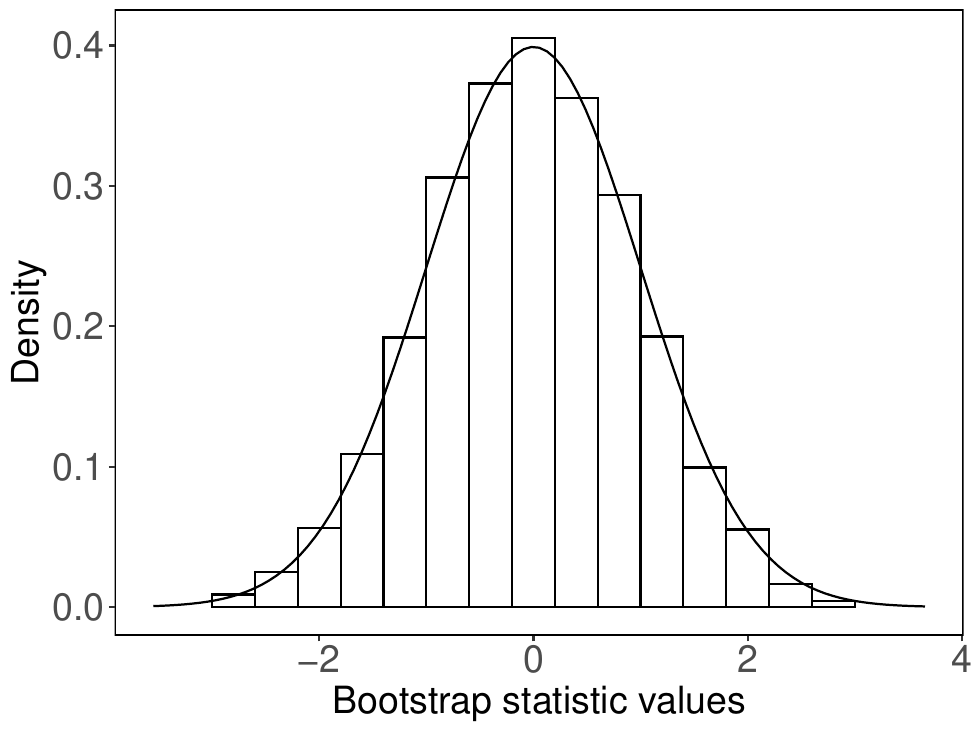} 
\caption{The histogram of $[\log(1-\hat{\psi}^2)-\hat{\delta}_\nu]/\hat{\sigma}_\nu$ values
from 10,000 bootstrap resamples and the density curve of the standard normal distribution.}
\label{fig3}
\end{figure}

Finally, we use the bootstrap with replacement to obtain the $95\%$ confidence interval for the coefficient of
multiple correlation $\psi$. 
To maintain consistency with the calculation of the asymptotic confidence interval, 
we first compute the confidence interval for $\log(1-\psi^2)$ and then transform it back to get
 the confidence interval for $\psi$. Based on 10,000 bootstrap replications, the $95\%$ confidence interval 
from the nonparametric bootstrap procedure is $(0.437,0.447)$, which is nearly the same as
the asymptotic confidence interval. When implementing the
bootstrap procedure, we also record the calculated values of $[\log(1-\hat{\psi}^2)-\hat{\delta}_\nu]/\hat{\sigma}_\nu$ 
from the 10,000 bootstrap resamples. The histogram of these values is shown in Figure
\ref{fig3}, together with the density curve of $N(0,1)$. From
the figure, it is obvious that the distribution of the statistic $[\log(1-\hat{\psi}^2)-\hat{\delta}_\nu]/\hat{\sigma}_\nu$ 
can be well approximated by $N(0,1)$, confirming the effectiveness of the normal approximation (\ref{equ11}) in this 
example. 

\section{Concluding Remarks}
\label{sec5}

The assessment of multiple correlation is an important problem with broad applications.
In this paper, we propose a new coefficient of multiple correlation that eliminates the need for specifying
the dependent variable. Under the assumption that data follow an independent component
model with independent but not necessarily components and finite fourth moments, we derive 
the asymptotic distribution of the sample coefficient and propose a bias-corrected estimator. 
It has been shown through theoretical justifications
and numerical studies that the proposed coefficient provides a useful tool for quantifying
the linear correlation among continuous variables.

However, there are still some issues with the proposed method that need to be addressed
in future research. For instance, the proposed coefficient relies on the Pearson-type correlation matrix
and thus it is mainly designed for measuring the linear correlation among continuous variables. 
To enhance versatility, future work could focus on developing a more general coefficient of 
multiple correlation which is capable of evaluating other types of multiple correlation, perhaps utilizing 
the Kendall's rank correlation matrix.
In addition, our theoretical results are derived under the assumption that data follow an independent 
component model, a popular choice in multiple correlation studies. Future efforts should be made to investigate 
the asymptotic distribution of the sample coefficient under alternative models, such as the elliptical model.
Furthermore, we assume that the components of the independent component model 
have finite fourth moments when studying the sampling distribution of the coefficient and the data dimension
is smaller than the sample size. In some applications, 
however, the data distribution could be heavy-tailed without a well defined fourth moment,
and the data could be high-dimensional in the sense that $p>n$. Subsequent research could explore theoretical extensions of the sampling
distribution to accommodate such scenarios. All
these issues need to be studied carefully in the future.

\end{document}